\documentclass[12pt,a4paper]{article}
\usepackage[dvips]{graphicx}

\newcommand{\kk}{$K^0$--$\bar{K}^0$}
\newcommand{\ek}{$\varepsilon_K$}
\newcommand{\bqbq}{$B_q$--$\bar{B}_q$}
\newcommand{\bsbs}{$B_s$--$\bar{B}_s$}
\newcommand{\bdbd}{$B_d$--$\bar{B}_d$}
\newcommand{\dmbq}{$\Delta m_q$}

\newcommand{\dmbsd}{$\Delta m_s/\Delta m_d$}
\newcommand{\bsg}{$b \to s\,\gamma$}
\newcommand{\meg}{$\mu \to e\,\gamma$}
\newcommand{\tmg}{$\tau \to \mu\,\gamma$}

\newcommand{\Bbsg}{${\rm B}(b \to s\,\gamma)$}
\newcommand{\Bmeg}{${\rm B}(\mu \to e\,\gamma)$}
\newcommand{\Btmg}{${\rm B}(\tau \to \mu\,\gamma)$}

\newcommand{\dlt}{$\delta_{13}$}

\begin{document}
\baselineskip 16pt

\begin{flushright}
\begin{tabular}{l}
KEK-TH-713 \\
September 2000
\end{tabular}
\end{flushright}

\vskip 5ex

\begin{center}

\lineskip 0.5ex
{\LARGE\bf
Neutrino Yukawa couplings and FCNC processes in $B$ decays in SUSY-GUT%
\footnote{
Talk given in
International Seminar Quarks-2000, May 14--21, 2000, Pushkin, Russia
and
XXXth International Conference on High Energy Physics, July 27--August
2, 2000, Osaka, Japan.
}
}

\vskip 3ex

{\large
Seungwon Baek, Toru Goto and Yasuhiro Okada
}

\vskip 1ex

{\it
Theory Group, KEK, Tsukuba, Ibaraki, 305-0801, Japan
}

\vskip 2ex

{\large
Ken-ichi Okumura
}

\vskip 1ex

{\it
Institute for Cosmic Ray Research, University of Tokyo,
Kashiwa, Chiba, 277-8582, Japan
}

\end{center}

\vskip 3ex

\begin{abstract}
Flavor changing neutral current and lepton flavor violating processes
are studied in the SU(5) SUSY-GUT with right-handed neutrino
supermultiplets.
Using input parameters motivated by neutrino oscillation, it is shown
that the time-dependent CP asymmetry of \bsg\ can be as large as 20\%.
We also show that the \bsbs\ mixing can be significantly different from
the standard model prediction.
\end{abstract}

\vskip 3ex

Effects of the physics beyond the standard model (SM) may appear in the
flavor physics in quarks and leptons.
One of indications is already given by the atmospheric and the solar
neutrino anomalies which are interpreted as evidences of neutrino
oscillation \cite{nuosc}.
A natural way to explain small neutrino masses is the see-saw mechanism
with very heavy right-handed neutrinos.
This scenario suggests the existence of new sources of flavor mixings
in the lepton sector at much higher energy scale than the electroweak
scale.

In this work \cite{Baek:2000sj}
we consider flavor changing neutral
current (FCNC) and lepton flavor violation (LFV) processes in the model
of a SU(5) supersymmetric (SUSY) grand unified theory (GUT) which
incorporates the see-saw mechanism for the neutrino masses.
In the SUSY model based on the minimal supergravity the mass
matrices of squarks and sleptons are flavor-blind at the Planck scale
$M_P$.
However renormalization effects due to Yukawa coupling constants of
quarks, leptons and neutrinos
induce flavor mixing in the squark/slepton mass matrices.
In the context of SUSY-GUT with right-handed neutrinos,
the flavor mixing related to the neutrino oscillation can provide
the mixing in the squark sector.
We show that due to the large mixing of the second and third generations
suggested by the atmospheric neutrino data, \bsbs\ mixing and the
CP asymmetry of the $B \to M_s\,\gamma$ process, where $M_s$ is a CP
eigenstate including the strange quark, have significant deviations from
the SM predictions.

The relevant part of the superpotential for the SU(5) SUSY GUT with
right-handed neutrino supermultiplets is given by
\begin{equation}
  W =
    \frac{1}{8}f_U^{ij} \Psi_{i} \Psi_{j} H_{5}
  + f_D^{ij} \Psi_{i} \Phi_{j} H_{\bar{5}}
  + f_N^{ij} N_{i} \Phi_{j} H_{5}
  + \frac{1}{2} M_{\nu}^{ij} N_{i} N_{j}
~,
\label{eq:SP}
\end{equation}
where $\Psi_{i}$, $\Phi_{i}$ and $N_{i}$ are ${\bf 10}$, $\bar{\bf 5}$
and ${\bf 1}$ representations of SU(5) gauge group.
$i,j=1,2,3$ are the generation indices.
$H_{5,\bar{5}}$ are Higgs superfields with ${\bf 5}$ and $\bar{\bf 5}$
representations.
$f_{U,D,N}$ are Yukawa coupling matrices and $M_\nu$ is the Majorana
mass matrix.
Below the GUT scale $M_G$ and the Majorana mass scale $M_R$ the
superpotential is written as
\begin{equation}
  W =
    \tilde{f}_U^{ij} Q_i U_j H_2
  + \tilde{f}_D^{ij} Q_i D_j H_1
  + \tilde{f}_L^{ij} E_i L_j H_1
  - \frac{1}{2} \kappa_{\nu}^{ij} (L_i H_2)(L_j H_2)
~,
\label{Wmssm}
\end{equation}
where $\kappa_{\nu}$ is obtained by integrating out $N_i$ at $M_R$.
The Yukawa coupling constants $\tilde{f}_U$, $\tilde{f}_D$
and $\tilde{f}_L$ are related to
$f_U$ and $f_D$ at $M_G$.
The masses and mixings of the quarks and leptons are determined
from the superpotential Eq.\ (\ref{Wmssm}) at the low energy scale.

As discussed above, the renormalization effects due to the Yukawa
coupling constants induce various FCNC/LFV effects from the
mismatch between the bases of quark/lepton and squark/slepton masses.
In particular the top Yukawa coupling constant is responsible for
the running of the $\tilde{q}_L$ and $\tilde{u}_R$ masses.
At the same time the $\tilde{e}_R$ mass matrix receives sizable
corrections between $M_P$ and $M_G$ scales and LFV processes are
induced \cite{LFV}.
In a similar way, if $f_N^{ij}$ is large, the $\tilde{l}_L$ and
$\tilde{d}_R$ mass matrices receive sizable flavor changing effects
due to the running between $M_P$ and $M_{G,R}$ scales \cite{nuR-LFV}.
These are sources of extra contributions to various FCNC/LFV processes.

We calculate the following observables:
the CP violation parameter in the \kk\ mixing \ek,
\bqbq\ mass splitting \dmbq\ ($q=d,s$),
the branching ratios of \bsg, \meg\ and \tmg,
and the amplitude of
the time-dependent CP asymmetry in the $B\to M_s\,\gamma$
process \cite{AGS-CHH},
which is written as
\begin{eqnarray}
   A_t &=&
   \frac{2{\rm Im}({\rm e}^{-i\theta_B} c_7 c'_7)}{|c_7|^2 + |c'_7|^2}
~,
\end{eqnarray}
where
$c_7$ and $c'_7$ are the Wilson coefficients in the effective
Lagrangian for the \bsg\ decay
$
{\cal L} = ( c_7 \bar{s} \sigma^{\mu\nu} b_R
          +c'_7 \bar{s} \sigma^{\mu\nu} b_L ) F_{\mu\nu} + {\rm H.c.}
$.
$\theta_B = \arg M_{12}(B_d)$ where
$M_{12}(B_d)$ is the \bdbd\ mixing amplitude.

We solved renormalization group equations (RGEs) for Yukawa coupling
matrices and the SUSY breaking parameters keeping all the flavor
mixings.
We specify neutrino parameters as well as the quark/lepton masses and
the Cabibbo-Kobayashi-Maskawa (CKM) matrix as follows.
The inputs from the neutrino oscillation are two mass differences and
the Maki-Nakagawa-Sakata (MNS) matrix.
In order to relate these inputs to $f_N$ and $M_\nu$, we work in the
basis where $\tilde{f}_L$ is diagonal and $f_N=\hat{f}_N V_L$
($\hat{f}_N$ is diagonal).
In this basis
$
\kappa_\nu = V_L^{\bf T}\hat{f}_N M_\nu^{-1} \hat{f}_N V_L
$
at the matching scale $M_R$.
Once we fix three neutrino masses, $V_{{\rm MNS}}$, $\hat{f}_N$ and
the unitary matrix $V_L$ we can obtain $M_\nu$.
Then using the GUT relation for Yukawa coupling constants, we calculate
all squark/slepton mass matrices through RGEs.
Note that $V_L$ essentially determines the flavor mixing in the
$\tilde{d}_R$ and $\tilde{l}_L$.

We consider the following parameter sets, corresponding to the
Mikheyev-Smirnov-Wolfenstein solutions for the solar neutrino
problem.
(i) small mixing angle solution:
$
  \sin^2 2\theta_{12} = 5.5\times 10^{-3}
$,
$
  m_{\nu} =
  ( 2.24,\, 3.16,\, 59.2 ) \times 10^{-3}\, {\rm eV}
$,
(ii) large mixing angle solution:
$
  \sin^2 2\theta_{12} = 1
$.
$
  m_{\nu} =
  ( 4.0,\, 5.83,\, 59.5 ) \times 10^{-3}\, {\rm eV}
$,
In both cases we take $\sin^2 2\theta_{23}=1$,
$\sin^2 2\theta_{13}=0$ and $M_{\nu}$ to be proportional to a unit
matrix with a diagonal element of $M_R = 4 \times 10^{14}\, {\rm GeV}$
so that $V_L = V_{{\rm MNS}}^{\dagger}$ at $M_R$.
Free parameters in the minimal supergravity model are the universal
scalar mass $m_0$, the unified gaugino mass $M_0$, the scalar
trilinear parameter $A_0$, the ratio of two vacuum expectation values
$\tan\beta$ and the sign of the Higgsino mass parameter $\mu$.
We take $\tan\beta=5$ and vary other SUSY parameters.
We also impose various constraints from SUSY particles search, the
measurement of \Bbsg\ \cite{bsgExp} and the search of \meg\ \cite{megExp}.

\begin{figure}[tb]
\begin{center}
\includegraphics[scale=0.5]{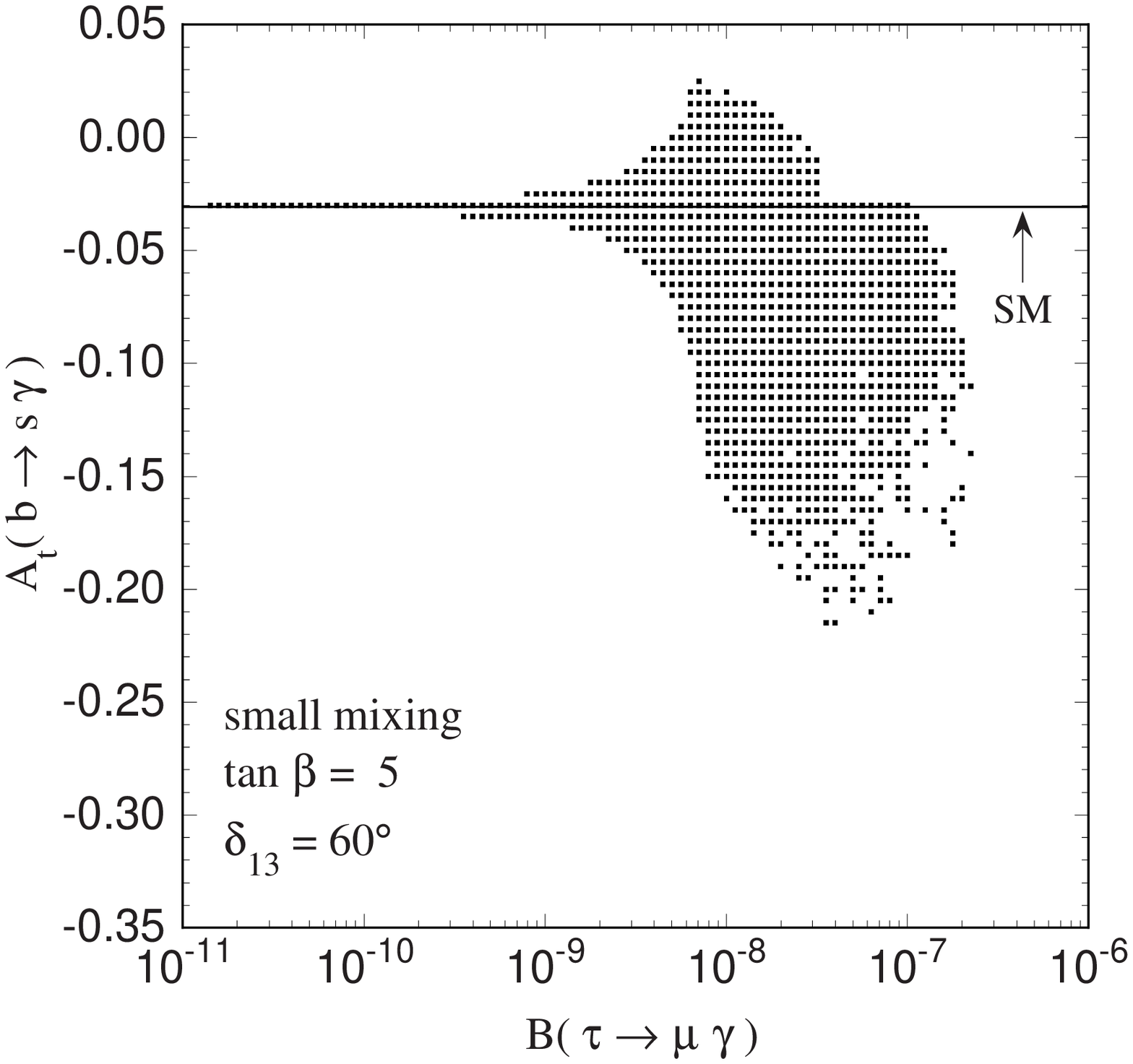}
\end{center}
\caption{
  Time-dependent CP asymmetry in $b\to s\gamma$ decay as a function of
  the branching ratio of $\tau\to\mu\gamma$.
}
\label{fig:Atbsg-taumug}
\end{figure}
Fig.~\ref{fig:Atbsg-taumug} shows $A_t$ as a function of \Btmg\ for case
(i) with fixed CKM parameters $|V_{ub}/V_{cb}|=0.08$ and
$\delta_{13}=60^{\circ}$ where \dlt\ is the phase parameter in the CKM
matrix.
We see that $|A_t|$ can be as large as 20\% when \Btmg\ is larger than
$10^{-8}$ level.
The large asymmetry arises because the renormalization effect of $f_N$
induces sizable contribution to $c'_7$ through gluino--$\tilde{d}_R$
loop diagrams.
Since this asymmetry is suppressed by a factor $m_s/m_b$ in the SM, a
sizable asymmetry is a clear signal of new physics beyond the SM.

\begin{figure}[tb]
\begin{center}
\includegraphics[scale=0.5]{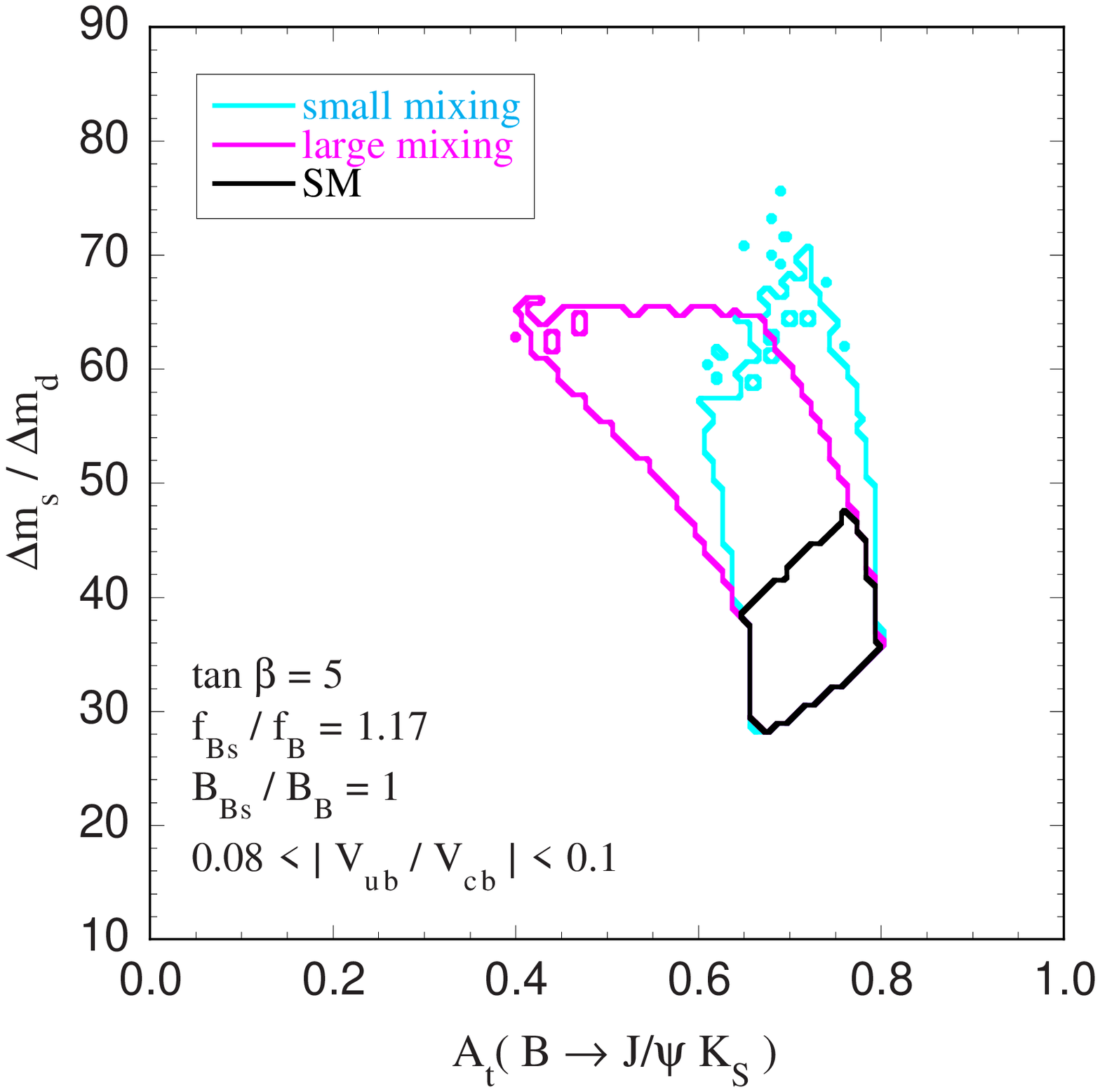}
\end{center}
\caption{
  Allowed regions in the space of \dmbsd\ and the CP asymmetry in $B\to
  J/\psi K_S$ decay.
}
\label{fig:dms-AtJKs}
\end{figure}
In Fig.~\ref{fig:dms-AtJKs} we show allowed regions in the space of
\dmbsd\ and the time-dependent CP asymmetry of $B\to J/\psi\,K_S$
for the case (i) and (ii).
Here $|V_{ub}|$ is varied within $0.08<|V_{ub}/V_{cb}|<0.1$ and \dlt\ is
scanned for the whole range.
For the case (i) we see that the deviation of $A_t(B\to J/\psi\,K_S)$
from the SM value is small while \dmbsd\ can differ from the SM
value by 40\%.
This pattern of deviation is understood as follows.
The new contributions to \ek\ and $M_{12}(B_d)$ are suppressed due to
the small 1-2 and 1-3 mixings in the neutrino sector so that the allowed
region of \dlt\ does not change much.
The deviation in \dmbsd\ comes from the SUSY contribution to
$M_{12}(B_s)$ induced by the large 2-3 mixing in the neutrino sector.
On the other hand we see a correlation between the deviations in the
case (ii).
Due to the large 1-2 mixing, \ek\ can be enhanced even after imposing
the \Bmeg\ constraint in this case.
Consequently the allowed range of \dlt\ by the constraint from \ek\
changes.
The region with large deviations in both \dmbsd\ and
$A_t(B\to J/\psi\,K_S)$ corresponds to a small \dlt\ region where the
constraint from \ek\ is satisfied by a large SUSY contribution.
This figure means the deviation from the SM may be seen in both cases
once \dmbsd\ and $A_t(B\to J/\psi\,K_S)$ are measured precisely.

In conclusion, we studied the effects of the neutrino Yukawa coupling
matrix on FCNC/LFV processes in the SU(5) SUSY-GUT with right-handed
neutrino supermultiplets.
It is shown that $A_t(B\to M_s\,\gamma)$ can be $\sim20$\% when the
\Btmg\ is about $10^{-7}$.
We also show that the \bsbs\ mixing can be significantly different from
the presently allowed range in the SM.
Since these signals provide quite different signatures compared to
the SM and the minimal supergravity model without GUT and right-handed
neutrino interactions, future experiments in $B$ physics and LFV can
give us important clues on the interactions at very high energy scale.

\end{document}